\documentclass[english,reprint,showpacs,prl,amsmath,amssymb,floatfix,superscriptaddress]{revtex4-1}
\usepackage[T1]{fontenc}
\usepackage[latin9]{inputenc}
\usepackage{color}
\usepackage{babel}
\usepackage{bm}
\usepackage{amsmath}
\usepackage{amssymb}
\usepackage{graphicx}
\providecommand{\U}[1]{\protect\rule{.1in}{.1in}}
\setcitestyle{numbers,square}
\begin{document}

\title{Voltage control of interface rare-earth magnetic moments}
\author{Alejandro~O. Leon$^{1}$, Adam B. Cahaya$^{1}$, and Gerrit E.~W. Bauer$^{1,2}$}
\affiliation{$^1$Institute for Materials Research, WPI-AIMR, and CSNR, Tohoku University, Sendai 980-8577, Japan}
\affiliation{$^2$4Zernike Institute for Advanced Materials, University of Groningen, 9747 AG Groningen, The Netherlands}
\date{\today }

\begin{abstract}
The large spin orbit interaction in rare earth atoms implies a strong coupling
between their charge and spin degrees of freedom. We formulate the coupling
between voltage and the local magnetic moments of rare earth atoms with
partially filled 4f shell at the interface between an insulator and a metal.
The rare earth-mediated torques allow power-efficient control of spintronic
devices by electric field-induced ferromagnetic resonance and magnetization switching.

\end{abstract}
\maketitle

\textbf{Introduction -} The power demand for magnetization control in magnetic
memories is an important design parameter. The power consumption of voltage
driven magnetization dynamics can be orders of magnitude lower than the one of
electric current induced dynamics~\cite{Nozaki2012}. In addition, electric
voltages are a more localized driving mechanism compared to magnetic
fields~\cite{Nozaki2012}. From the experimental point of view, magnetization
reversal~\cite{Shiota2009,Kanai2012} and ferromagnetic
resonance~\cite{Nozaki2012,Zhu2012} driven by electric voltages have been
achieved. In those studies, transition-metal films are capped with an
insulating barrier that prevents the electric current flow. The main mechanism
to couple voltage and magnetization is the control of the perpendicular
magnetic anisotropy~\cite{Suzuki2011}. Other realizations of the magnetization
manipulation by electric fields were conducted in (Ga,Mn)As
semiconductor~\cite{DChiba2008} and materials with magnetoelectric
properties~\cite{Gerhard2010,Yamada2011,Sekine2016}.

The spin-charge coupling that lies beyond the observed phenomena have been
modeled by the Rashba~\cite{RashbaOriginal,RashbaReviewDuine} and
Dzyaloshinskii-Moriya
interactions~\cite{RashbaReviewDuine,Dzyaloshinskii,Moriya}, whose origin is
the relativistic magnetic field induced by \textit{linear momentum} of the
electron in a transverse electric field. On the other hand, the spin-orbit
interaction in central fields of single atoms can best be expressed in terms
of the effective magnetic field generated by orbital \textit{angular
momentum}. Here we focus on local magnetic moments in condensed matter system
for which the second picture of the spin-orbit interaction is the best
starting point.

Local magnetic moments in solids are formed by partially filled $\mathrm{3d}$
and $\mathrm{4f}$ shells of transition metals and rare earths, respectively.
The former are relatively light and their spin dynamics are dominated by the
exchange interaction, with correction by the crystal fields. Rare earths (RE),
on the other hand, have their magnetic sub shell shielded by outer shells,
which decreases the effect of crystal-fields and allows the electrons to orbit
almost freely in the central Coulomb field of the ionic core with large
nuclear charges. The spin-orbit interaction (SOI) of RE is therefore large and
free-atomic like. Since SOI couples the electric and magnetic degrees of
freedom, we may expect significant effects of electric fields on the RE
magnetization dynamics.

Here we study the voltage-driven dynamics of rare earths at the interface
between a magnetic insulator (or bad conductor) and a metal. When one of the
layers is magnetic, the presence of RE at the interface strongly couples the
magnetization to an applied static and dynamic voltage by the local the
spin-orbit interaction. Electric fields, applied by high-frequency signal
generators for example, are constant inside an insulator but nearly vanish in
a metal. The large spatial gradients of the electric field at the interface
re-normalize the RE electrostatic interactions with neighboring atoms
(crystal-fields), and appear as a voltage modulated magnetic anisotropy and
the associated magnetization torque that we derive in the following in more detail.

\textbf{Magnetism of rare earth ions -} In the Russell-Saunders
scheme~\cite{BookJensen} the total spin ($\mathbf{S}$) and orbital
($\mathbf{L}$) momenta are the sum of the single electron momenta of the
$\mathrm{4f}$-orbitals $\mathbf{S}=\sum_{j}\mathbf{s}_{j}$ and $\mathbf{L}%
=\sum_{j}\mathbf{l}_{j}$. The spin-orbit coupling reads $H_{\mathrm{SOI}%
}=\Lambda\mathbf{S}\cdot\mathbf{L}$, and the coupling parameter $\Lambda$ is
positive (negative) for less (more) than half-filled sub
shell~\cite{BookJensen,BookBlundell}. The total angular momentum vector
$\mathbf{J}=\mathbf{S}+\mathbf{L}$ and the angular part of the eigenfunctions
can be written as $|\Psi\rangle=|S,L,J,J_{z}\rangle$, where the quantum
numbers are governed by $\mathbf{S}^{2}|\Psi\rangle=\hbar^{2}S(S+1)|\Psi
\rangle$, $\mathbf{L}^{2}|\Psi\rangle=\hbar^{2}L(L+1)|\Psi\rangle$,
$\mathbf{J}^{2}|\Psi\rangle=\hbar^{2}J(J+1)|\Psi\rangle$, $\hat{J}_{z}%
|\Psi\rangle=\hbar J_{z}|\Psi\rangle$, $\hat{J}_{z}$ is the $z$-component of
the vector $\mathbf{J,}$ and $\hbar$ is Planck's constant divided by $2\pi$.
The lowest-energy state of RE ions as governed by Hund's
rules~\cite{BookJensen} are listed in Table~\ref{TableQ2}. The Wigner-Eckart
theorem ensures that within this ground state manifold the angular momenta are
collinear, viz. $\mathbf{S}=(g_{J}-1)\mathbf{J}$ and $\mathbf{L}%
=(2-g_{J})\mathbf{J}$ in terms of the Land\'{e} g-factor $g_{J}$. Furthermore,
for constant $(S,L,J)$ the orbital symmetry axis and the spin vector move
rigidly together, implying a strong spin-charge
coupling~\cite{BookSkomski1,BookSkomski2}.\begin{table}[b]
\begin{center}%
\begin{tabular}
[c]{|c|c|c|c|c|c|c|}\hline
Ion & $4f^{n}$ & S & L & J & Shape & $Q_{2}/a_{0}^{2}$\\\hline
Ce$^{3+}$ & $4f^{1}$ & $\frac{1}{2}$ & 3 & $\frac{5}{2}$ & Oblate &
-0.686\\\hline
Pr$^{3+}$ & $4f^{2}$ & 1 & 5 & 4 & Oblate & -0.639\\\hline
Nd$^{3+}$ & $4f^{3}$ & $\frac{3}{2}$ & 6 & $\frac{9}{2}$ & Oblate &
-0.232\\\hline
Pm$^{3+}$ & $4f^{4}$ & 2 & 6 & 4 & Prolate & 0.202\\\hline
Sm$^{3+}$ & $4f^{5}$ & $\frac{5}{2}$ & 5 & $\frac{5}{2}$ & Prolate &
0.364\\\hline
Eu$^{3+}$ & $4f^{6}$ & 3 & 3 & 0 & - & -\\\hline
Gd$^{3+}$ & $4f^{7}$ & $\frac{7}{2}$ & 0 & $\frac{7}{2}$ & Spherical &
0\\\hline
Tb$^{3+}$ & $4f^{8}$ & 3 & 3 & 6 & Oblate & -0.505\\\hline
Dy$^{3+}$ & $4f^{9}$ & $\frac{5}{2}$ & 5 & $\frac{15}{2}$ & Oblate &
-0.484\\\hline
Ho$^{3+}$ & $4f^{10}$ & 2 & 6 & 8 & Oblate & -0.185\\\hline
Er$^{3+}$ & $4f^{11}$ & $\frac{3}{2}$ & 6 & $\frac{15}{2}$ & Prolate &
0.178\\\hline
Tm$^{3+}$ & $4f^{12}$ & 1 & 5 & 6 & Prolate & 0.427\\\hline
Yb$^{3+}$ & $4f^{13}$ & $\frac{1}{2}$ & 3 & $\frac{7}{2}$ & Prolate &
0.409\\\hline
\end{tabular}
\end{center}
\caption{Ground state $(S,L,J)$ based on Hund's rules and shape of the $4f$
ground state electron density~\cite{BookSkomski1}. $Q_{2}$ is the quadrupole
moment calculated using the Wigner-Eckart theorem for the state $J_{z}=J$, and
$a_{0}=0.53$ \AA \ is the Bohr radius. The Wigner-Eckart theorem cannot be
applied to Eu because $\mathbf{J}=0$.}%
\label{TableQ2}%
\end{table}The electron density of a partially filled $\mathrm{4f}$ sub shell
can be written as
\begin{equation}
n_{\mathrm{4f}}(\mathbf{r})=\sum_{m_{l}=-3}^{3}|R_{\mathrm{4f}}(r)Y_{3}%
^{m_{l}}(\mathbf{\hat{r}})|^{2}\left(  f_{m_{l}\uparrow}+f_{m_{l}\downarrow
}\right)  ,
\end{equation}
where $\mathbf{r}=r\mathbf{\hat{r}}$ is the position vector in spherical
coordinates, $R_{\mathrm{4f}}(r)$ the radial part of the $\mathrm{4f}$
atomic-like wave function, and the spherical harmonics $Y_{3}^{m_{l}%
}(\mathbf{\hat{r}})$ describe the angular dependence. $f_{m_{l},m_{s}}$ is the
occupation number of the single electron state with magnetic quantum numbers
of orbital $m_{l}$ and spin $m_{s}$ angular momenta. The density
$n_{\mathrm{4f}}$ is normalized to the number of electrons in the
$\mathrm{4f}$ shell $N_{\mathrm{4f}}=\int n_{\mathrm{4f}}(\mathbf{r}%
,t)d\mathbf{r}$. The typical $\mathrm{4f}$ radius, $\langle r\rangle\sim0.5$
\AA , is much smaller than typical inter atomic distances, $R\sim3$%
\thinspace\AA , which motivates the multipole
expansion~\cite{BookSkomski1,BookSkomski2}
\begin{equation}
n_{\mathrm{4f}}(\mathbf{r})\approx\frac{|R_{\mathrm{4f}}(r)|^{2}}{4\pi
}\left\{  N_{\mathrm{4f}}+\frac{5Q_{2}}{4\langle r^{2}\rangle}\left[  3\left(
\mathbf{m}\cdot\mathbf{\hat{r}}\right)  ^{2}-1\right]  \right\}
,\label{EqN4fQuadrupolar}%
\end{equation}
where $Q_{2}\equiv\int\left(  3z^{2}-r^{2}\right)  n_{\mathrm{4f}}%
(\mathbf{r})d\mathbf{r}$ is the quadrupole moment listed in Table
\ref{TableQ2}. $\mathbf{m}=-\mathbf{J}/|\mathbf{J}|$ is the unit magnetization
vector that at equilibrium is taken to be $\mathbf{e}_{z}$ but in an excited
state may depend on time. The unit position vector in spherical coordinates is
$\mathbf{\hat{r}}=\sin\theta\left[  \mathbf{e}_{x}\cos\phi+\mathbf{e}_{y}%
\sin\phi\right]  +\mathbf{e}_{z}\cos\theta$, where $\{\mathbf{e}%
_{x},\mathbf{e}_{y},\mathbf{e}_{z}\}$ are the unit vectors along the Cartesian
axes. For $Q_{2}>0$ $\left(  Q_{2}<0\right)  $ the envelope function of the
electron density is a pancake or cigar-like\ (oblate or prolate) ellipsoid, respectively.

A local magnetic ion interacts weakly with static electric fields,
$\mathbf{E}=-\nabla V$, where $V$ is the voltage or potential energy of a
positive probe charge. To leading order, the ions experience the electrostatic
energy~\cite{SkomskiArticle}
\begin{equation}
\langle\psi|-e\sum_{i=1}^{N_{\mathrm{4f}}}V(\mathbf{r}_{i})|\psi\rangle=-e\int
d^{3}rV(\mathbf{r})n_{\mathrm{4f}}(\mathbf{r}), \label{EqMeanH}%
\end{equation}
where $-e$ is the electron charge, and $-eV(\mathbf{r}_{i})$ is the potential
energy of the $i$-th electron. $|\psi\rangle$ is the $\mathrm{4f}$
many-electron wave function in the ground state. Again, the leading order in a
multipole expansion of the crystal field around the origin $\mathbf{r}=0$ can
be parameterized by a quadrupolar term $A_{2}^{(0)}$
\begin{equation}
eV(\mathbf{r})=-A_{2}^{(0)}r^{2}\left(  3\cos^{2}\theta-1\right)  .
\label{EqVoltage}%
\end{equation}
Inserting Eqs.~(\ref{EqN4fQuadrupolar}) and~(\ref{EqVoltage}) into
Eq.~(\ref{EqMeanH}), we arrive at a Hamiltonian that depends on the
magnetization direction as
\begin{equation}
H_{\mathrm{ani}}=\frac{3}{2}Q_{2}A_{2}^{(0)}m_{z}^{2}. \label{EqEAnisoGeneral}%
\end{equation}
The crystal symmetry orients here the easy ($Q_{2}A_{2}^{(0)}<0$) or hard
($Q_{2}A_{2}^{(0)}>0$) magnetic axis along the $z$ direction. This
\textit{crystal field}~ energy accounts for the single rare-earth ion magnetic
anisotropy. The parameter $A_{0}^{(2)}$ can be calculated by first principles
or to fit to experiments. Typical values are: $A_{0}^{(2)}=300$\thinspace
$\mathrm{K}a_{0}^{-2}$ for (RE)$_{2}$Fe$_{14}$B, $A_{0}^{(2)}=34$\thinspace
K$a_{0}^{-2}$ for (RE)$_{2}$Fe$_{17}$, and $A_{0}^{(2)}=-$358 K $a_{0}^{-2}$
for (RE)$_{2}$Fe$_{17}$N$_{3}$~\cite{BookSkomski2}, where $a_{0}=0.53$
\AA \ is the Bohr radius. The origin of the strong magnetic anisotropy of REs
is their large spin-orbit interactions. On the other hand, for $3d$ transition
metal moments, the anisotropy is usually very small, except at interfaces,
where the orbital motions are partially unquenched. In such cases, the
anisotropy emerges as the consequence of SOI, the quadrupolar shape of
electric potentials at the interface~\cite{Suzuki2011}, the hybridization of
orbitals and change in the orbitals occupation.

At an interface between materials with different work functions the symmetry
is reduced by potential steps. The electric field exhibits spatial gradients
due to charge accumulation immediately at the interface that result in a step
potential. An external voltage difference $\Delta V$ drops only over the
insulator, but is constant in the metal (when the ferromagnet is a bad
conductor the effects are weaker but still exist). The electric field
therefore depends on position in the immediate proximity of the interface.
This dependence electric field gradients can interact with the 4f sub shell as
an effective crystal field.

\begin{figure}[h]
\includegraphics[width=\columnwidth]{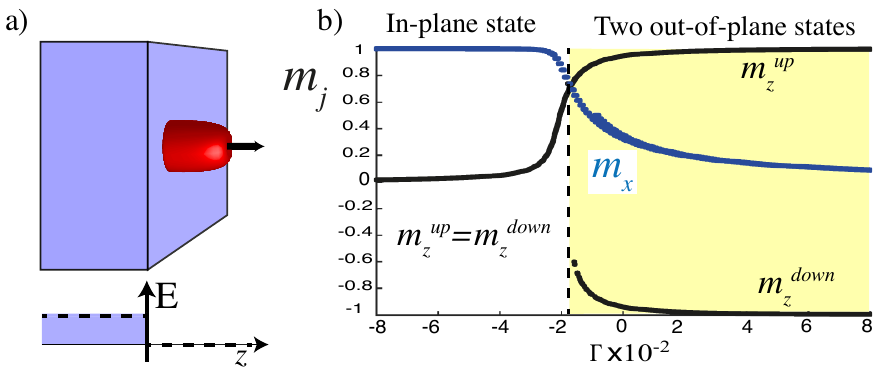} \centering\caption{Spin-charge
coupling for an interface local magnetic moment. a) Electric field at an
interface between an insulator (constant electric field) and a metal
(vanishing electric field). The magnetic dipole and charge quadrupole at the
interface are strongly coupled. b) Ground state magnetization directions
$\mathbf{m}=(m_{x},m_{y},m_{z})$ as a function of interface electric field
with coupling parameter $\Gamma$ [Eq. (\ref{Gamma})] and a magnetic field
tilted from the z-axis by an angle $\varphi\sim6^{\circ}$. The system switches
from a perpendicular easy axis to an easy-plane configuration for
$\Gamma<0.018.$}%
\label{FigCoupling}%
\end{figure}

\textbf{Voltage coupling at interfaces -} Let us focus on a magnetic insulator
film with thickness $L_{F}$. At the surface, the insulator exposes
$n_{\mathrm{RE}}$ rare earth moments per unit of area. Inside the insulator,
the electric field is approximately constant, $\mathbf{E}(z<0)=\mathbf{e}%
_{z}\Delta V/L_{F}$, while it vanishes in the metal $\mathbf{E}(z>0)=0$, see
Fig.~\ref{FigCoupling}a). Using Eq.~(\ref{EqMeanH}), the electric energy of a
magnetic moment at the origin is then
\begin{equation}
H_{e}=H_{0}-\frac{15}{64}\frac{e\Delta V}{L_{F}}Q_{2}\frac{\langle r\rangle
}{\langle r^{2}\rangle}m_{z}^{2}\label{EqEAniI}%
\end{equation}
where $H_{0}=5eE_{0}Q_{2}\langle r\rangle/(64\langle r^{2}\rangle)+e(\Delta
V/L_{F})N_{\text{\textrm{4f}}}\langle r\rangle/4$ does not depend on the
magnetization and $\langle r^{n}\rangle\equiv N_{\mathrm{4f}}^{-1}\int
r^{n}n_{\text{\textrm{4f}}}(\mathbf{r})d\mathbf{r}$. In the
Supplement~\cite{Supplemental} different approaches to formulate the coupling
yield expressions similar to Eq. (\ref{EqEAniI}). For $\langle r^{2}%
\rangle^{1/2}\sim\langle r\rangle\sim0.5$\thinspace\AA \ the coupling energy
per unit area at equilibrium ($m_{z}^{2}=1$)
\begin{equation}
\left\vert n_{\mathrm{RE}}(H_{e}-H_{0})\frac{L_{F}}{\Delta V}\right\vert
=750\frac{\mathrm{fJ}}{\mathrm{Vm}}\frac{Q_{2}}{10^{-3}\mathrm{nm}^{2}}%
\frac{n_{\mathrm{RE}}}{\mathrm{nm}^{-2}}%
\end{equation}
is one order of magnitude larger than the corresponding coupling in transition
metals~\cite{TransitionMetal2011,TransitionMetal2012}. For electric fields
$\Delta V/L_{F}\sim10$ mV/nm $=100$ kV/cm, the surface energy density becomes
$(H_{e}-H_{0})n_{\mathrm{RE}}\approx7.5\times10^{-3}$\thinspace erg/cm$^{2}%
=7.5\,\mathrm{\mu}$J/m$^{2}$.

The step field model can also be applied to \textit{non-magnetic
insulators}$|$\textit{transition-metal ferromagnets} (such as Fe, Co, and Ni,
or their alloys) with RE ions at the interface that are antiferromagnetically
coupled to the magnetic order \cite{REDoping,RECapping} and facilitate a large
coupling of the magnetization to electric fields. Good insulators, such as
MgO, can endure very large electric fields (of the order of $300$ mV/nm, in
FeCo$|$MgO, Ref.~\cite{TransitionMetal2011}, for example). Thus, MgO based
magnetic tunnel junctions with rare earth doping or dusting are promising
devices to study and apply electric field-induced modulations of the
magnetization configuration.

In magnetic materials, local angular momenta are strongly locked by the
exchange interaction. When a sufficiently strong static magnetic field
$\mathbf{B}$ is applied, the macrospin model is valid, i.e. the magnetization
$\mathbf{M}$ is constant in space. The total magnetic energy $H_{M}$ per unit
area then reads
\begin{equation}
\frac{H_{M}}{\mu_{0}M_{s}^{2}L_{F}}=-\mathbf{m}\cdot\mathbf{h}-\frac{\beta
_{x}}{2}m_{x}^{2}+\frac{\beta_{z}-\Gamma}{2}m_{z}^{2}.\label{EqEnergy}%
\end{equation}
The first term on the (dimensionless) right-hand side with $\mathbf{h}%
=\mathbf{B/}\left(  \mu_{0}M_{s}\right)  $ is the Zeeman energy and $M_{s}$
the saturation magnetization. The parameters $\beta_{x}$ ($\beta_{z}$) account
for the in-plane (out of plane) magnetic anisotropy in the absence of applied
electric fields, $\Delta V=0$. The dimensionless coupling parameter $\Gamma$
measures the relative strength of the electrostatic coupling $\sim
n_{\mathrm{RE}}e\Delta VQ_{2}/L_{F}$ that should be compared with magnetic
anisotropies. $\Gamma\sim0.06$ with the following parameters representative
for a rare earth iron garnet thin film such as Tm$_{3}$Fe$_{5}$O$_{12}$
\begin{equation}
\Gamma=0.06\frac{n_{\mathrm{RE}}}{1/\mathrm{nm}^{2}}\left(  \frac
{10^{5}\mathrm{A/m}}{M_{s}}\right)  ^{2}\left(  \frac{10\,\mathrm{nm}}{L_{F}%
}\right)  ^{2}\frac{\Delta V}{0.1\,\mathrm{V}}\frac{Q_{2}}{10^{-3}%
\mathrm{nm}^{2}}\label{Gamma}%
\end{equation}
Since the $M_{s}$ of 8 nm thick Tm$_{3}$Fe$_{5}$O$_{12}$~\cite{Tang2016} is at
room temperature about 10 times smaller than that of even a subnanometer FeCo
film~\cite{TransitionMetal2011}, the coupling strength $\Gamma$ is 10 times
larger for magnetic insulators for the same applied electric field without the
need for additional tunnel barriers. Intraband transition and electric
breakdown is of no concern as long as $eE_{0}\ll\epsilon_{\mathrm{gap}}%
^{2}/(\epsilon_{F}a)$, where $\epsilon_{\mathrm{gap}}$ is the band gap,
$\epsilon_{F}$ is the Fermi level in the metal and $a$ the lattice
constant~\cite{AshdroftMerminBook}. Using $\epsilon_{F}\sim2$ eV, and the
gap/lattice constant for yttrium iron garnet (YIG)~\cite{YIGExp1,YIGExp2}
$\epsilon_{\mathrm{gap}}\sim2.85$ $\mathrm{eV}$/$a=1.2$ nm, we estimate
$E_{0}\ll2$ $\mathrm{V}$/nm to be safe. The coupling strength $\Gamma$
decreases $\sim L_{F}^{-2}$ for a given voltage, so much can be gained by
choosing an insulator with a large gap and breakdown voltage that permits
working with thin layers.

Figure \ref{FigCoupling}b) shows the stable magnetizations that minimize of
the energy~(\ref{EqEnergy}) in the presence of a magnetic field $\mathbf{h}%
=h[\mathbf{e}_{x}\cos\varphi+\mathbf{e}_{z}\sin\varphi]$ that is tilted by an
angle $\varphi.$ The parameter are $h=0.01$, $\varphi=5.72^{\circ}$, $\beta_{x}=0$,
$\beta_{z}=-0.03$. The application of a constant voltage allows the transition
from the easy axis (right zone) to the easy plane (left zone) configuration.

The electric field effects in transition metal devices as well as one proposed
here, derive from the same type of magnetic anisotropy, although the
microscopic coupling mechanism is different. The phenomenology of electric
field-induced precessional dynamics as observed in transition metal
systems~\cite{ReviewVCM} does not differ from the one we expect for RE
systems. The advantage of interface REs is the lower power consumption and the
possibility of using a wider range of materials including magnetic insulators,
such as YIG. The magnetization dynamics is described by the
Landau-Lifshitz-Gilbert equation,
\begin{equation}
\dot{\mathbf{m}}=-\gamma\mathbf{m}\times\mathbf{h}_{\mathrm{eff}}%
+\alpha\mathbf{m}\times\dot{\mathbf{m}}, \label{EqLLG}%
\end{equation}
where $\alpha$ is Gilbert damping constant, $\gamma>0$ is the (modulus of the)
gyromagnetic ratio, $\dot{\mathbf{m}}$ the temporal derivative of $\mathbf{m}%
$, and the effective magnetic field $\mathbf{h}_{\mathrm{eff}}$ satisfying
\begin{equation}
\frac{\mathbf{h}_{\mathrm{eff}}}{\mu_{0}M_{s}}\equiv\frac{1}{\mu_{0}M_{s}%
^{2}L_{F}}\frac{\partial H_{M}}{\partial\mathbf{m}}=\mathbf{h}+\beta_{x}%
m_{x}\mathbf{e}_{x}+\left[  \Gamma-\beta_{z}\right]  m_{z}\mathbf{e}_{z}.
\label{EqHeffEinducedSP}%
\end{equation}
The magnetic torque exerted by the electric field is proportional to
$-\mathbf{m}\times\gamma m_{z}\mathbf{e}_{z}$.

\begin{figure}[h]
\includegraphics[width=\columnwidth]{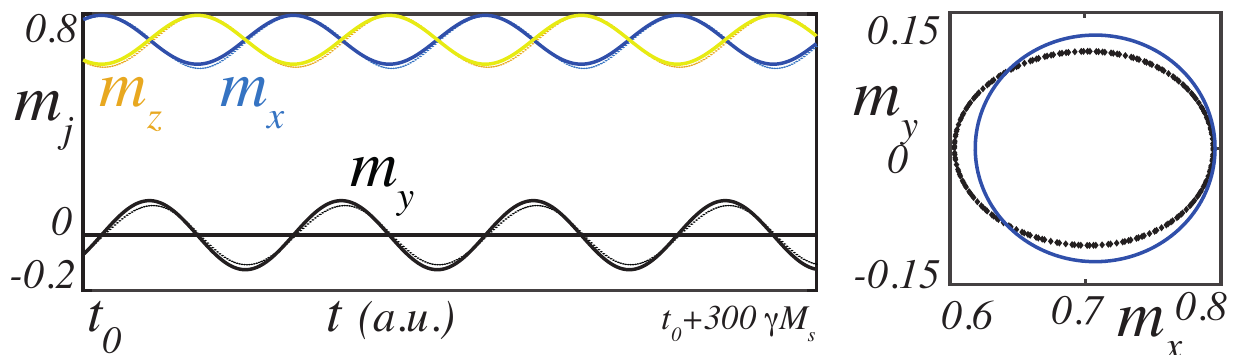} \centering\caption{ (a)
Magnetization dynamics induced by time-varying voltages. Comparison between
analytic (solid line) and (b) numeric (dot) precessional (FMR) solutions,
obtained for $\Gamma=\Gamma_{0}\cos(\Omega t)$, $\Gamma_{0}=0.01$, $\beta
_{x}=\beta_{z}=0$, $\Omega=0.08$, $h=0.1$, $\varphi=45^{\circ}$, and $\alpha=0.005$. }%
\label{FigComp}%
\end{figure}

\textbf{Ferromagnetic resonance - }We now turn to an ac electric field that
modulates the coupling $\Gamma=\Gamma_{0}\cos(\Omega t)$, with frequency
$\Omega$ close to the ferromagnetic resonance (GHz). Since the electric field
is normal to thin metallic films $<100$ nm, the induced Oersted-like magnetic
field and associated power are negligibly small. In linear response the
model~(\ref{EqLLG}) can be solved analytically for $\beta_{x}=\beta_{z}=0$.
The polar coordinate system is spanned by the unit vectors $\mathbf{e}%
_{1}=\mathbf{e}_{x}\cos\left(  \varphi\right)  +\mathbf{e}_{z}\sin\varphi$,
$\mathbf{e}_{2}=-\mathbf{e}_{x}\sin\left(  \varphi\right)  +\mathbf{e}_{z}%
\cos\varphi$, and $\mathbf{e}_{3}=-\mathbf{e}_{y}$. At equilibrium state
$\mathbf{m}_{\mathrm{eq}}=\mathbf{e}_{1}$ along the applied magnetic field.
Around the equilibrium state, the magnetization is $\mathbf{m}=\mathbf{e}%
_{1}+\delta\mathbf{m}$, where $\delta\mathbf{m}=\delta m_{2}\mathbf{e}%
_{2}+\delta m_{3}\mathbf{e}_{3}$ is the deviation from $\mathbf{m}%
_{\mathrm{eq}}$, with $\left\vert \delta\mathbf{m}\right\vert \ll1$ and
$\delta\mathbf{m}\cdot\mathbf{m}_{\mathrm{eq}}=0$. To leading order in the
coupling ($\Gamma_{0}$) and dissipation ($\alpha$), the effective field is
$(\mu_{0}M_{s})^{-1}\mathbf{h}_{\mathrm{eff}}=h\mathbf{e}_{1}+\Gamma
\cos(\Omega t)\sin(\varphi)\left[  \mathbf{e}_{1}\sin(\varphi)+\mathbf{e}%
_{2}\cos\varphi\right]  $ and
\[
\frac{\delta\dot{\mathbf{m}}}{\omega_{M}}=\mathbf{e}_{1}\times\left[
h\delta\mathbf{m}+\alpha\frac{\delta\dot{\mathbf{m}}}{\omega_{M}}-\frac
{\Gamma_{0}}{2}\cos(\Omega t)\sin(2\varphi)\mathbf{e}_{2}\right]  ,
\]
where $\omega_{M}\equiv\gamma\mu_{0}M_{s}$. The effective ac magnetic field
$\mathbf{B}_{\mathrm{ac}}=\mu_{0}M_{s}\Gamma_{0}\sin(2\varphi)\cos(\Omega
t)\mathbf{e}_{z}/2$. Then
\begin{align}
\delta\mathbf{m} &  =(\Gamma_{0}/4)\sin(2\varphi)\chi^{\prime}(\Omega)\left(
\cos(\Omega t)\mathbf{e}_{2}+\sin(\Omega t)\mathbf{e}_{3}\right)  \nonumber\\
&  +(\Gamma_{0}/4)\sin(2\varphi)\chi^{\prime\prime}(\Omega)\left(  \sin(\Omega
t)\mathbf{e}_{2}-\cos(\Omega t)\mathbf{e}_{3}\right)  ,\nonumber
\end{align}
where $\chi^{\prime}$ and $\chi^{\prime\prime}$ are the real and imaginary
parts of the dynamics susceptibility
\[
\chi(\omega)\equiv\frac{\omega_{M}\left(  \omega_{0}-\omega\right)  }{\left(
\omega_{0}-\omega\right)  ^{2}+\omega^{2}\alpha^{2}}+i\frac{\omega_{M}%
\alpha\omega}{\left(  \omega_{0}-\omega\right)  ^{2}+\omega^{2}\alpha^{2}},
\]
and the natural frequency is $\omega_{0}\equiv\omega_{M}h=\gamma\mu_{0}M_{s}%
h$. Figure~\ref{FigComp} illustrates $\delta\mathbf{m}(t)$ (continuous lines)
together with the numeric solution (dots). We see that a large oscillation
cone $\left\vert \delta\mathbf{m}\right\vert \sim0.15$ can be achieved by a
relatively low voltage for the aforementioned parameter values and $\Gamma
_{0}=0.01$ (or $\Delta V/L_{F}\sim1.6$ $\mathrm{mV/nm}$).

\begin{figure}[h]
\includegraphics[width=\columnwidth]{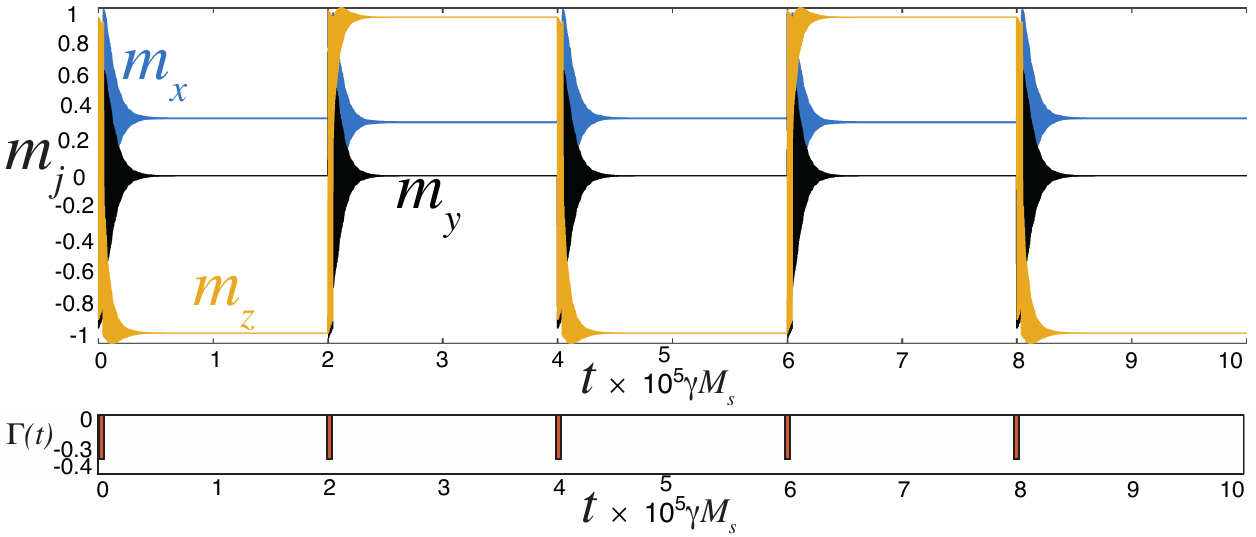} \centering\caption{
Precessional switching for an easy axis perpendicular magnet ($\beta
_{z}=-0.03$) induced by a voltage box train. Up pannel shows the magnetization components, while the low panel shows the box consisting of a negative
voltage with $\Gamma=-0.03$ for $\Delta t=$ $4000\gamma M_{s}$ ($\sim$ 1 ns)
followed by $\Gamma=0$. This signal is repeated all $2\times10^{5}\gamma
M_{s}$ ($\sim$25 ns). Other parameters are $\beta_{x}=0$, $h=0.01$, $\varphi
=5.72^{\circ}$, and $\alpha=0.005$.}%
\label{FigCompII}%
\end{figure}

\textbf{Magnetization switching - }Magnetic reversal in tunnel junctions is
the key process in magnetic random access memories. An applied voltage can
reduce the energy barrier for magnetic field and current-induced switching or
directly trigger the magnetization reversal~\cite{ReviewVCM}. The latter
effect is illustrated by Fig.~\ref{FigCompII} assuming perpendicular
magnetization (for in-plane magnetization, see
Ref.~\cite{Shiota2009,Kanai2012}). An equilibrium magnetization along $z$
[either an $\emph{up}$ or $\emph{down}$ state in the right zone of
Fig.~\ref{FigCoupling}b)] is excited by a step-like voltage pulse into large
damped precessions around the in-plane equilibrium [left zone of
Fig.~\ref{FigCoupling}b)]. When the voltage is turned off again at the right
time, the magnetization can be fully reverted. The switching is observed with
large tolerance in the pulses duration between the pico and nano second scales. In the
simulation of Fig.~\ref{FigCompII}, the pulse duration is around 1 ns, while
the application of subsequent pulses toggles the magnetization direction
faithfully.

\textbf{Conclusions and remarks -} We report voltage-modulated magnetic
anisotropies and magnetization dynamics of rare-earth magnetic moments at
insulator$|$metal bi-layer interfaces. An applied voltage generates
inhomogenous electric fields at interfaces with large conductivity mismatch
that couple efficiently to rare earth ions with non-spherical electron
distributions, which is usually the case when the shell is not half or
completely filled. The dynamics of the charge and spin distributions are
locked by the spin-orbit interaction. The voltage can then rigidly precess the
charge and spin distributions of the entire $\mathrm{4f}$ sub shell via a
stronger and direct coupling to the spin than in transition metals. Adding
rare-earth impurities to insulator$|$metal bi-layers can be used to
efficiently switch the magnetization and induce ferromagnetic resonance.
Future applications may include rare earth-dusted magnetic insulator$|$normal
metal interfaces, such as YIG$|$Pt, that can efficiently convert an ac voltage
into a spin current by spin-pumping.

\textit{Acknowledgments.-} We acknowledge the financial support from JSPS
KAKENHI Grants Nos. 25247056, 25220910, and 26103006 and JSPS Fellowship for
Young Scientists No. JP15J02585. We profited from initial research by Dr.
Mojtaba Rahimi.

\onecolumngrid \clearpage \begin{center} \textbf{\large  Supplemental Material to Voltage control of interface rare-earth magnetic moments} \end{center}
\setcounter{equation}{0} \setcounter{figure}{0} \setcounter{table}{0} \setcounter{page}{1} \makeatletter \renewcommand{\theequation}{S\arabic{equation}} \renewcommand{\thefigure}{S\arabic{figure}}

This supplemental material shows alternative derivations of the coupling
between voltage and rare earth magnetic moments. Details on the numerical
simulation of the voltage-induced dynamics are also presented.

\section{Rare earth impurities at a metal interface}

\begin{figure}[b]
\includegraphics[width=11 cm]{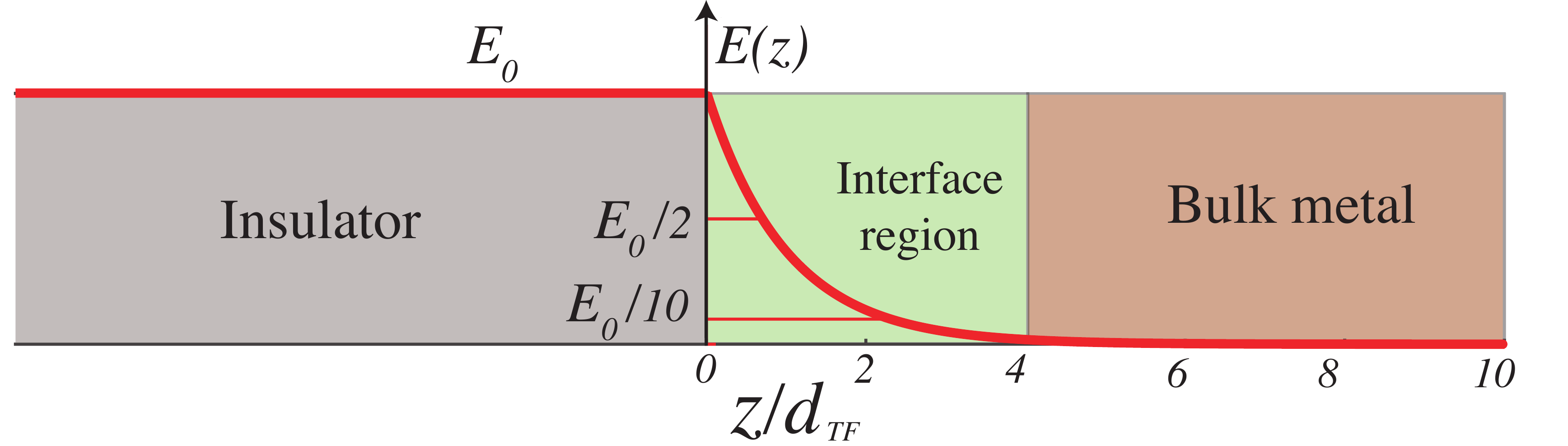} \centering\caption{(Color online)
Interface screening. The electric field is finite inside the insulator and
vanishes deeper inside the metal. At the interface, the charge accumulation
generates gradients in the electric field and potential. Only inside the metal
and very close to the interface (\emph{interface region}) the rare earth ions
is affected by an applied electric voltage.}%
\label{FigScreening}%
\end{figure}

Here we describe the insulator$|$magnetic metal bi-layer in terms of a
screening-induced crystal field shift, which is a Thomas-Fermi-like
justification of the step model used in the main text. The electric field in a
metal normal to the interface to an insulator (in the $x-y$ plane) reads in a
local screening model
\begin{equation}
\mathbf{E}(z)=E_{0}e^{-z/d_{\mathrm{TF}}}\mathbf{e}_{z},\label{EqETM}%
\end{equation}
where $E_{0}$ is the electric field in the insulator and $d{_{\mathrm{FT}}}$
is the Thomas-Fermi screening length. Figure~\ref{FigScreening} illustrates
the electric field profile. For a non-magnetic metal $d{_{\mathrm{FT}}}%
\equiv\lbrack\epsilon_{0}/(e^{2}g)]^{1/2}$, where $g$ is the conduction
electron density of states at Fermi level and $\epsilon_{0}$ is the
permittivity of free space. In ferromagnetic metals, $d{_{\mathrm{TF}}}$ can
be obtained using Poisson equation and a Stoner model (see
Ref.~\cite{FMScreening}). For elemental metals the screening length of the
order of an \AA . For example, for Fe(bcc), $d{_{\mathrm{TF}}}=1.3$
\AA ~\cite{FMScreening}. In close proximity of a given atom we have then a
modified \textquotedblleft crystal field\textquotedblright\ that can be
expressed by the leading terms of a Taylor expansion
\begin{align}
\mathbf{E}(z) &  \approx E_{0}\mathbf{e}_{z}+\left(  \frac{\partial
E}{\partial z}\right)  _{0}z\mathbf{e}_{z}=-\mathbf{e}_{z}\frac{\partial
}{\partial z}\left[  -E_{0}z-\frac{1}{2}\left(  \frac{\partial E}{\partial
z}\right)  _{0}z^{2}\right]  ,\nonumber\\
&  =-\mathbf{e}_{z}\frac{\partial}{\partial z}\left[  -E_{0}z-\frac{1}%
{6}\left(  \frac{\partial E}{\partial z}\right)  _{0}\left(  3z^{2}%
-r^{2}\right)  -\frac{1}{6}\left(  \frac{\partial E}{\partial z}\right)
_{0}r^{2}\right]  .
\end{align}
The potential near the interface has dipolar [$-E_{0}z$], an isotropic
[$-(1/6)\left(  \partial_{z}E\right)  _{0}r^{2}$] and uni-axial [quadrupolar:
$-(1/6)\left(  \partial_{z}E\right)  _{0}(3z^{2}-r^{2})$] contributions. The
latter can be estimated from Eq.~(\ref{EqETM}) close to $z=0$,
\begin{equation}
V(\mathbf{r})=\frac{r^{2}E_{0}}{6d{_{\mathrm{TF}}}}\left(  3\cos^{2}%
\theta-1\right)  ,
\end{equation}
which corresponds to the magnetic anisotropy energy
\begin{equation}
H_{\mathrm{ani},M}=-\frac{eQ_{2}}{4d{_{\mathrm{TF}}}}E_{0}m_{z}^{2}%
,\label{EqEnergyAniE}%
\end{equation}
where $m_{z}$ is the unit magnetization component along the interface normal
($m_{z}=\mathbf{m}\cdot\mathbf{e_{z}}$), and $Q_{2}\equiv\int\left(
3z^{2}-r^{2}\right)  n_{\mathrm{4f}}(\mathbf{r})d\mathbf{r}$ is quadrupolar
moment of the rare earth ion. For typical values $d_{TF}\sim10^{-1}$ nm and
$Q_{2}\sim10^{-3}$ nm$^{2}$, this energy is of the same order of magnitude as
the one obtained for the simple step model in the main text [Eq. (6)].

\section{Torque derivation using Newtonian mechanics}

The strong spin-charge coupling in rare earth atoms implies also the locking
between the atom angular momentum and the $4f$ sub shell mass distribution.
This property allows us to derive the torque exerted by voltages using the
simple approach of the Newtonian mechanics, which we present below for the
non-specialist reader.

The force acting on a charge element $dQ=-en_{\mathrm{4f}}(\mathbf{r})d^{3}r$
of volume element $d\mathbf{r}$ at $\mathbf{r}$ is
\begin{equation}
d\mathbf{F}(\mathbf{r})=-en_{\mathrm{4f}}(\mathbf{r})\mathbf{E}(\mathbf{r}%
)d\mathbf{r},
\end{equation}
The corresponding (mechanical) torque is
\begin{equation}
\mathbf{T_{m}}=\int\mathbf{r}\times d\mathbf{F}(\mathbf{r})=-e\int
n_{\mathrm{4f}}(\mathbf{r})\mathbf{r}\times\mathbf{E}(\mathbf{r})d\mathbf{r},
\end{equation}
and it acts on a non-spherical electron distribution as parameterzed by the
quadrupole moment $Q_{2}$ and oriented along the unit vector $\mathbf{m}$.
Expanding the electric field in a Taylor series near the origin,
$\mathbf{E}(\mathbf{r})=[E_{0}+z(\partial E/\partial z)_{0}]\mathbf{e_{z}}$
\begin{align}
\mathbf{T_{m}}  &  =\frac{e\langle r^{2}\rangle}{4\pi}\left(  \frac{\partial
E}{\partial z}\right)  _{0}\mathbf{e}_{z}\times\int_{0}^{\pi}d\theta
\sin(\theta)\cos(\theta)\int_{0}^{2\pi}d\phi\mathbf{\hat{r}}\left[
N_{4f}+\frac{5Q_{2}}{4\langle r^{2}\rangle}\left(  3\left[  \mathbf{m}%
(t)\cdot\frac{\mathbf{r}}{r}\right]  ^{2}-1\right)  \right]  ,\\
&  =-\frac{eQ_{2}}{2}\left(  \frac{\partial E}{\partial z}\right)  _{0}\left(
\mathbf{m}\cdot\mathbf{e}_{z}\right)  \mathbf{m}\times\mathbf{e}%
_{z}=\mathbf{m}\times\frac{\delta}{\delta\mathbf{m}}\left[  -\frac{eQ_{2}}%
{4}\left(  \frac{\partial E}{\partial z}\right)  _{0}\left(  \frac{M_{z}%
}{M_{s}}\right)  ^{2}\right]  .
\end{align}
The orbital and magnetic (classical) momenta are related by the gyromagnetic
ratio $-\gamma$ (where $\gamma>0$). The mechanical torque $\mathbf{T_{m}}$ is
therefore proportional to the magnetic torque $\mathbf{T}:$
\[
\mathbf{T}=\gamma\frac{eQ_{2}}{2}\left(  \frac{\partial E}{\partial z}\right)
_{0}\left(  \mathbf{m}\cdot\mathbf{e}_{z}\right)  \mathbf{m}\times
\mathbf{e}_{z}%
\]
which enters the Landau-Lifshitz-Gilbert equation.

\section{Numerical simulations}

The dimensionless Landau-Lifshitz-Gilbert equation, $\dot{\mathbf{m}%
}=-\mathbf{m}\times\mathbf{h}_{\mathrm{eff}}+\alpha\mathbf{m}\times
\dot{\mathbf{m}}$, decomposed in Cartesian coordinates and written in the
Landau-Lifshitz form:
\begin{align}
(1+\alpha^{2})\frac{dm_{x}}{dt} &  =\alpha m_{x}\left(  m_{z}^{2}\left[
\beta_{x}+\beta_{z}-\Gamma\right]  +\beta_{x}m_{y}^{2}\right)  +(\beta
_{z}-\Gamma)m_{y}m_{z}\nonumber\\
&  -h\sin(\varphi)(\alpha m_{x}m_{z}+m_{y})+\alpha h\cos(\varphi)\left(
m_{y}^{2}+m_{z}^{2}\right)  ,\nonumber\\
(1+\alpha^{2})\frac{dm_{y}}{dt} &  =-\beta_{x}m_{x}(\alpha m_{x}m_{y}%
+m_{z})+m_{z}\left[  \beta_{z}-\Gamma\right]  (\alpha m_{y}m_{z}%
-m_{x})\nonumber\\
&  +h\sin(\varphi)(m_{x}-\alpha m_{y}m_{z})-h\cos(\varphi)(\alpha m_{x}%
m_{y}+m_{z}),\nonumber\\
(1+\alpha^{2})\frac{dm_{z}}{dt} &  =-\alpha m_{z}\left(  \beta_{x}m_{x}%
^{2}+\left[  \beta_{z}-\Gamma\right]  \left(  m_{x}^{2}+m_{y}^{2}\right)
\right)  +\beta_{x}m_{x}m_{y}\nonumber\\
&  +\alpha h\sin(\varphi)\left(  m_{x}^{2}+m_{y}^{2}\right)  +h\cos
(\varphi)(m_{y}-\alpha m_{x}m_{z}).\label{EqLL}%
\end{align}
which is rendered dimensionless by measuring time in units of $\left(  \gamma
M_{s}\right)  ^{-1}$. We solve set of Eqs.~(\ref{EqLL}) by using a fifth order
Runge-Kutta scheme based on Ref.~\cite{NumericalRecipes}. A time step of
$\Delta t=0.01$ [in units of $(\gamma M_{s})^{-1}$]\textit{ }is sufficiently
small for accurate results. We monitored norm conservation, by requiring
$\left\vert 1-(m_{x}^{2}+m_{y}^{2}+m_{z}^{2})^{1/2}\right\vert \leq10^{-6}$.
The graphs in the main text were plotted after integrating the equation of
motion for a transient time of order $t_{0}=10^{5}$. Unless mentioned
explicitly, the resulting dynamics does not depend on the initial condition.

\subsection{Stationary states of the LLG equation}

The stationary state ($\dot{\mathbf{m}}=0$) in the presence of a constant
applied voltage as parameterized by $\Gamma$ should satisfy $\mathbf{m}%
\times\mathbf{h_{\mathrm{eff}}}=0$, or
\begin{equation}
\mathbf{m}=\left(  \frac{h\cos(\varphi)}{\lambda-\beta_{x}},0,\frac
{h\sin(\varphi)}{\lambda+\beta_{z}-\Gamma}\right)  ,
\end{equation}
where the $\lambda$ is obtained from the normalization condition $\left\vert
\mathbf{m}\right\vert =1$. By integrating the set of equations~(\ref{EqLL})
for different initial conditions, we obtain the shifted equilibrium states, as
shown in Fig. 1 of the main text.

\subsection{Ferromagnetic resonance}

\begin{figure}[t]
! \includegraphics[width=.7\columnwidth]{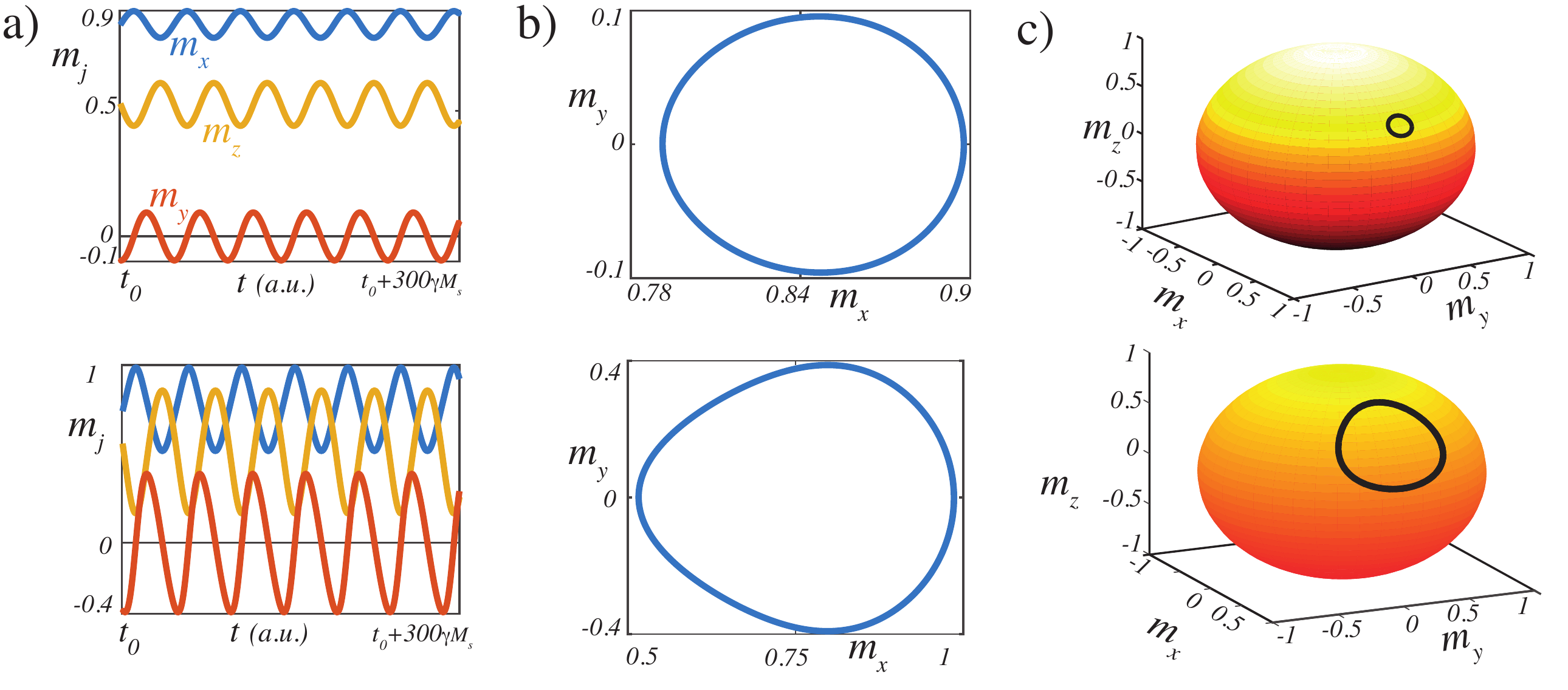}
\centering\caption{(Color online) Voltage induced ferromagnetic resonance. The
up and down panels show the ac electric field- induced steady-state
magnetization dynamics calculated for coupling parameters $\Gamma_{0}=0.01$
and $\Gamma_{0}=0.05$, respectively. a) Cartesian components as function of
time. b) Phase portrait of the variables $(m_{x},m_{y})$. c) Precession cone
on the unit sphere $\mathbf{m}^{2}=1$. Other parameters are $\Omega=0.08$,
$h=0.1$, $\phi=45^{\circ}$, $\beta_{x}=0.001$, $\beta_{z}=0.05$ and $\alpha=0.005$.
}%
\label{FigTraj}%
\end{figure}

\begin{figure}[b]
\includegraphics[width=.7\columnwidth]{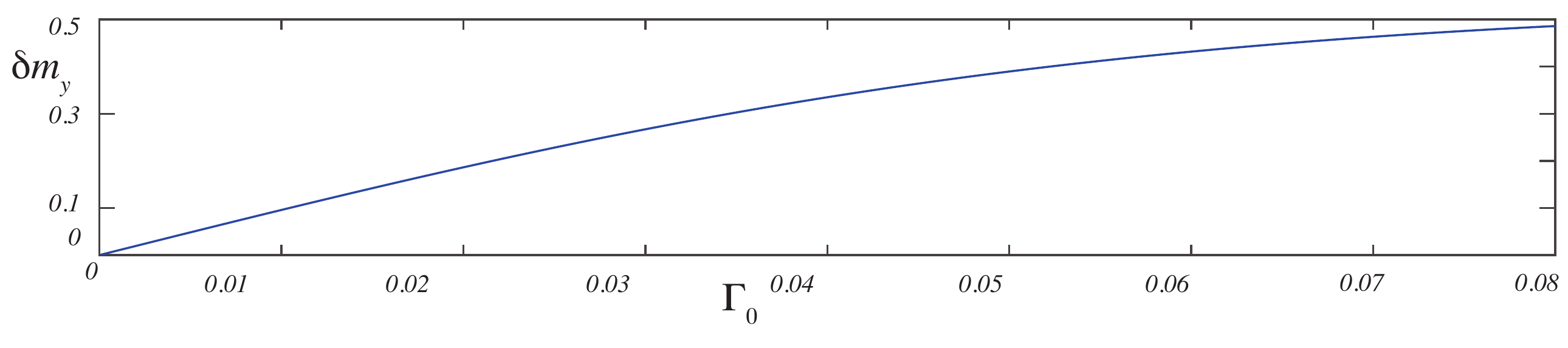}
\centering\caption{Magnetization oscillation amplitude at resonance as
function of the spin-charge coupling parameter $\Gamma_{0}$.}%
\label{FigAmpl}%
\end{figure}

As oscillating voltage $\Gamma(t)=\Gamma_{0}\cos(\Omega t)$ leads to
resonance, as illustrated in Figure~\ref{FigTraj} for two different voltage
amplitudes, and Fig.~\ref{FigAmpl} shows the precession cone as function of
the angle. Large oscillations ($|\delta\mathbf{m}|$ of about 10\% of $M_{s}$)
can be achieved for relatively low values of the spin-charge coupling
parameter ($\Gamma_{0}\sim0.01$)\textit{ } 


\begin{references}
\bibitem{Dipolar}
As a consequence of strong electric fields,  the electron polarize (Stark effect). This is a bulk effect. Furthermore, the energy of the Stark effect
scales~\cite{Suzuki2011} as $E_{0}^{2}$, and then this is a much
smaller effect, that we disregard here.
\end{references}
\end{document}